# Stable massless scalar polarization of $f(R)$ gravity

Xin-Dong Du[*] and Peng-Cheng Li[†]

School of Physics and Optoelectronics, South China University of Technology,

Guangzhou 510641, People's Republic of China

**Abstract**

Polarization is a prominent feature of gravitational wave observations and can be used to distinguish between different modified gravity theories. Compared to General Relativity, $f(R)$ gravity exhibits an additional polarization originating from a scalar field, which is a combination of the longitudinal and breathing modes. When the scalar mass of $f(R)$ is zero, the mixed mode will reduce to a pure breathing mode with the disappearance of the longitudinal mode. However, this reducing seems to be disallowed because a positive scalar mass is often required to maintain the stability of the cosmological perturbation. In fact, the massless scalar case can provide a stable perturbation, but more detailed constraints need to be considered. For the completeness of the polarization analysis, we explore the possibility that there are stable massless scalar polarizations in viable dark energy $f(R)$ models. We find that the existence of stable massless scalar polarization depends on the structure of $f(R)$ model and can be used to distinguish different models in $f(R)$ gravity.

## 1. Introduction

All kinds of cosmological observations have indicated that our Universe is expanding at an accelerating rate [1-3]. To explain accelerating expansion, there are two main approaches: introducing dark energy and modifying Einstein's General Relativity [4, 5]. The latter corresponds to the modified gravity theories, which include $f(R)$ theory [6, 7], Brans-Dicke theory [8], Horndeski theory [9], and general scalar-tensor theory [10] and so on. In addition to explaining accelerating expansion, modified gravity theories have also been used to explain inflation [11, 12]. In $f(R)$ gravity, General Relativity is modified by replacing the Ricci scalar $R$ in the Einstein-Hilbert action with an arbitrary function of $R$, denoted as $f(R)$. When the field equations are derived from the action in $f(R)$ gravity, two formalisms need to be distinguished. The first is the metric formalism, where the connections are set to be metric dependent. The second is the Palatini formalism [13], where the metric and the connections are assumed to be independent of each other. In this paper, viable $f(R)$ dark energy models will be studied in the metric formalism.

[*] 3537494784@qq.com

[†] Corresponding author: pchli2021@scut.edu.cn

Gravitational waves have been successfully observed [14, 15], and an increasing number of gravitational wave events are expected to be detected as various gravitational wave detectors continue to develop [16-18]. Therefore, it is possible to test different models of modified gravity according to the observations of gravitational waves [19-21]. One of the most significant properties of gravitational waves is the polarization. General Relativity has two independent polarizations coming from tensor fields, while general four-dimensional modified gravity theories have up to six independent polarizations [22]. As a result, polarization can be applied to test modified gravity theories [22-24]. And pulsar timing arrays can be used as an observation tool to detect different polarizations [25, 26]. There are some methods for polarization analysis, mainly including using geodesic deviation [27, 28], Newman-Penrose formalism [22, 29], extended Newman-Penrose formalism [30] and gauge invariants [31-33]. In the following section, geodesic deviation will be used to analyze polarizations.

For $f(R)$ gravity, its propagating degree of freedom is three, and it has an additional polarization coming from a scalar field besides the two same tensor polarizations as General Relativity [28]. The scalar polarization of $f(R)$ is a combination of the longitudinal and the breathing modes, and the longitudinal mode will disappear when the scalar mass of $f(R)$ is zero [28]. In other words, the massless scalar polarization of $f(R)$ is a pure breathing mode rather than a mixed mode [34]. However, a positive value of scalar mass is often required to maintain the stability of the cosmological perturbation [35, 36], so that the massless case seems to disappear. In fact, the massless case can provide a stable perturbation, but more detailed constraints need to be considered [37]. In order to build a complete polarization analysis for $f(R)$ gravity, it is necessary for us to give these detailed constraints and explore the possibility of the existence of stable massless scalar polarization in $f(R)$ gravity again.

It is interesting to note that the massless scalar polarization is allowed to appear in the Brans-Dicke theory [33, 38], but $f(R)$ gravity, which is largely equivalent to the former [39, 40], rarely has the massless scalar polarization. This inconsistency also prompts us to study the massless scalar polarization of $f(R)$ gravity. Although relevant examples have been researched in [34, 41], some problems remain unresolved. In [34], a special power law $f(R)$ model is considered, but it lacks the main features of viable $f(R)$ models. In [41], the massless scalar polarization is given by directly setting the scalar mass to zero, but more detailed constraints for a stable perturbation are not involved, so that the de Sitter point in the effective potential picture might be an inflection point or a local maximum point instead of an expected local minimum point.

Our paper is organized as follows: in Section 2, we introduce the polarizations of $f(R)$ gravity and three viable $f(R)$ models for dark energy. Particularly, the expression of scalar mass of $f(R)$ is given. In Section 3, some necessary constraints for stable massless scalar polarization are shown, and they come from the requirements of cosmology and effective potential. In Section 4, the updated constraints are used to retest viable $f(R)$ models whether they can reflect stable massless scalar polarizations. In Section 5, discussions and conclusions are made. In this paper, we use the geometrized units $G = c = 1$ and the signature $(-+++)$.

## 2. Polarizations and $f(R)$ gravity

In this section, we are going to introduce the gravitational wave polarizations of $f(R)$ gravity and the specific viable $f(R)$ models individually. The polarization analysis will focus on a particular model $f(R) = R + \alpha R^2$ ($\alpha > 0$) for a simple process, because the polarization results of general $f(R)$ model are similar to that of the particular model but require more complex calculations.

### 2.1 Wave solutions

The action for $f(R)$ gravity is given by:

$$S = \frac{1}{2\kappa} \int d^4x \sqrt{-g} f(R) + S_m, \tag{1}$$

where $\kappa = 8\pi$, $R$ is the Ricci scalar, $f(R)$ is an arbitrary function of $R$, and $S_m$ is the matter action. To study gravitational waves in the absence of matter, we set $S_m = 0$ for the vacuum. From the action, the perturbation equations for scalar and tensor can be obtained. In [36], the perturbation equation for scalar is given by perturbing a background curvature:

$$\left(\Box_g - m^2\right) \delta R = 0, \tag{2}$$

where $\Box_g = g^{\mu\nu}\nabla_\mu\nabla_\nu$ is the D'Alembert operator for the background metric $g_{\mu\nu}$, and the scalar mass $m$ is defined as:

$$m^2 = \frac{1}{3}\left(\frac{f'(R_d)}{f''(R_d)} - R_d\right), \quad (3)$$

where $f'(R) = df(R)/dR$, $f''(R_d) = df'(R)/dR$ and $R_d$ is a constant background curvature of de Sitter stage ($R_d$ is generally positive and could be viewed as zero in the limiting case). According to [42], to keep de Sitter stage static, $R_d$ should meet the following condition:

$$f'(R_d)R_d - 2f(R_d) = 0. \quad (4)$$

From here on, the subsequent discussion is based on the particular model $f(R) = R + \alpha R^2$ ($\alpha > 0$) for the simplicity of process [28]. According to Eq. (4), we have $R_d = 0$ for the particular model. Therefore, the background metric $g_{\mu\nu}$ becomes Minkowski metric $\eta_{\mu\nu}$, and the metric perturbation over it is:

$$\eta_{\mu\nu} + h_{\mu\nu}, \quad (5)$$

where $|h_{\mu\nu}| \ll 1$. In this case, the d'Alembert operator becomes $\Box_g = \Box_\eta = \eta^{\mu\nu}\nabla_\mu\nabla_\nu$. A new tensor $\bar{h}_{\mu\nu}$ is introduced:

$$\bar{h}_{\mu\nu} = h_{\mu\nu} - \frac{1}{2}\eta_{\mu\nu}h - 2\alpha\eta_{\mu\nu}\delta R, \quad (6)$$

where $h = \eta^{\mu\nu}h_{\mu\nu}$. In [28] [43], the perturbation equation for tensor is yielded:

$$\Box_\eta \bar{h}_{\mu\nu} = 0, \quad (7)$$

which is similar to the perturbation equation of General Relativity. Besides, general $f(R)$ model will also lead to the same result, see [36] for more details.

Based on the perturbation equations Eq. (2) and Eq. (7), their solutions for the plane wave traveling along the $z$ direction can be written as:

$$\delta R = \phi e^{ip_\mu x^\mu}, \quad (8)$$

$$\bar{h}_{\mu\nu} = \epsilon_{\mu\nu} e^{ikq_\mu x^\mu}, \quad (9)$$

where $\eta_{\mu\nu}p^\mu p^\nu = -m^2$, $p^\mu = (\Omega, 0, 0, \sqrt{\Omega^2 - m^2})$ and $q^\mu = (\omega, 0, 0, \omega)$ [28, 43]. The wave speed of $\delta R$ is $v = \sqrt{\Omega^2 - m^2}/\Omega$, and the wave speed of $\bar{h}_{\mu\nu}$ is the light speed $c = 1$. Further Eq. (8) and Eq. (9) can be rewritten as:

$$\delta R(vt - z), \quad (10)$$

$$\bar{h}_{\mu\nu} = \begin{pmatrix} 0 & 0 & 0 & 0 \\ 0 & \epsilon_+ & \epsilon_\times & 0 \\ 0 & \epsilon_\times & -\epsilon_+ & 0 \\ 0 & 0 & 0 & 0 \end{pmatrix} \cos(\omega(t-z)), \tag{11}$$

where $\delta R$ is expressed as the function of $vt - z$, $\epsilon_+$ and $\epsilon_\times$ are nonzero constants, the matrix of $\bar{h}_{\mu\nu}$ is given by applying the transverse traceless gauge conditions, and all phases have been ignored here.

## 2.2 Polarization modes

In the following part, we are going to bring the wave solutions into the geodesic deviation equations to show their polarization states. Considering the condition $\eta^{\mu\nu}\bar{h}_{\mu\nu} = 0$ [28], Eq. (6) can give the total metric perturbation $h_{\mu\nu}$:

$$h_{\mu\nu} = \bar{h}_{\mu\nu} - 2\alpha\eta_{\mu\nu}\delta R, \tag{12}$$

where $\bar{h}_{\mu\nu}$ stands for the tensor perturbation and $-2\alpha\eta_{\mu\nu}\delta R$ stands for the scalar perturbation. Because tensor and scalar perturbations are decoupled, we are allowed to calculate their polarization modes separately, and the total polarizations of $f(R)$ gravity will be their superposition.

For a wave traveling along the $z$ direction in a local proper reference frame, the deviation vectors $S^i$ between two adjacent geodesics are described by the geodesic deviation equations:

$$\frac{d^2 S^i}{dt^2} = -R^i_{0j0} S^j, \tag{13}$$

where $i, j = 1, 2, 3$, $S^i = (x, y, z)$, and $R^i_{0j0}$ are so-called "electric" components of the Riemann tensor. Under the first-order perturbation of the metric, $R^i_{0j0}$ become:

$$R^i_{0j0} = \eta_{ii} R^i_{0j0} = R_{i0j0} = \frac{1}{2}(\partial_i \partial_0 h_{0j} + \partial_0 \partial_j h_{i0} - \partial_0 \partial_0 h_{ij} - \partial_i \partial_j h_{00}). \tag{14}$$

Based on Eq. (10) and Eq. (11), bring Eq. (12) into Eq. (14) to obtain:

$$R_{i0j0} = -\frac{1}{2}(\partial_0 \partial_0 \bar{h}_{ij}) + \alpha(\eta_{ij}\partial_0\partial_0\delta R - \partial_i\partial_j\delta R). \tag{15}$$

And the above equation can be divided into tensor and scalar two parts.

For the tensor part, we can take $R_{i0j0} = -\frac{1}{2}(\partial_0 \partial_0 \bar{h}_{ij})$ and Eq. (11) into Eq. (13) to get:

$$\ddot{x} = -\frac{\omega^2}{2}\epsilon_+ \cos(\omega(t-z))x - \frac{\omega^2}{2}\epsilon_\times \cos(\omega(t-z))y, \tag{16}$$

$$\ddot{y} = -\frac{\omega^2}{2}\epsilon_\times \cos(\omega(t-z))x + \frac{\omega^2}{2}\epsilon_+ \cos(\omega(t-z))y, \tag{17}$$

where the terms containing $\epsilon_+$ represent the plus mode in gravitational wave polarizations, and the terms containing $\epsilon_\times$ represent the cross mode. Because $\epsilon_+$ and $\epsilon_\times$ are independent, the tensor polarizations of $f(R)$ gravity share two degrees of freedom, as the same as General Relativity.

For the scalar part, we can take $R_{i0j0} = \alpha(\eta_{ij}\partial_0\partial_0\delta R - \partial_i\partial_j\delta R)$ and Eq. (10) into Eq. (13) to get:

$$\ddot{x} = -\alpha\partial_0\partial_0\delta R x, \tag{18}$$

$$\ddot{y} = -\alpha\partial_0\partial_0\delta R y, \tag{19}$$

$$\ddot{z} = \alpha\left(\frac{m^2}{\Omega^2}\right)\partial_3\partial_3\delta R z, \tag{20}$$

where $m^2 = 1/(6\alpha)$ for the particular model $f(R) = R + \alpha R^2$. Here Eq. (20) is obtained by applying $\partial_0\partial_0\delta R = v^2 \partial_3\partial_3 \delta R$ and $v = \sqrt{\Omega^2 - m^2}/\Omega$. Eqs. (18) and (19) represent the breathing mode, while Eq. (20) represents the longitudinal mode. Because Eqs. (18), (19) and (20) are all controlled by the parameter $\alpha$, the scalar polarizations of $f(R)$ gravity share one degree of freedom. It means that the scalar polarizations of $f(R)$ should be a combination of the breathing and the longitudinal modes [28]. And the mixed mode will reduce to a pure breathing mode when the scalar mass $m$ of $f(R)$ (defined in Eq. (3)) vanishes. Although the above discussions are based on the particular model $f(R) = R + \alpha R^2$ over the Minkowski background, some similar results have been given for general $f(R)$ gravity over a de Sitter background [33], and even for other extended gravity theories which includes $f(R)$ gravity [32, 33].

## 2.3 Specific models of $f(R)$ gravity

Various $f(R)$ models are proposed mainly to explain accelerating expansion and inflation [11, 12, 44]. Here we are more concerned about the case of accelerating expansion. In order to explore the possibility of massless scalar polarization in $f(R)$ gravity, we are going to discuss several viable $f(R)$ dark energy models including Hu-Sawicki [45], Starobinsky [46] and Gogoi-Dev [41]:

$$f_1(R) = R - \alpha_1 R_c \frac{\left(\frac{R}{R_c}\right)^{2\beta_1}}{\left(\frac{R}{R_c}\right)^{2\beta_1} + 1} \quad \text{(Hu-Sawicki)}, \tag{21}$$

$$f_2(R) = R - \alpha_2 R_c \left[1 - \left(1 + \left(\frac{R}{R_c}\right)^2\right)^{-\beta_2}\right] \quad \text{(Starobinsky)}, \quad (22)$$

$$f_3(R) = R - \frac{\alpha_3}{\pi} R_c \text{arccot}\left(\left(\frac{R}{R_c}\right)^{-2}\right) - \beta_3 R_c \left(1 - e^{-\frac{R}{R_c}}\right) \quad \text{(Gogoi-Dev)}, \quad (23)$$

where $\text{arccot}(x)$ is the inverse function of $\cot(x)$, $\alpha_i$ and $\beta_i$ ($i = 1, 2, 3$) are positive dimensionless parameters, while $R_c > 0$ is a parameter in unit of curvature and roughly corresponds to the order of present Ricci scalar [7]. These above $f(R)$ models can replace dark energy to explain accelerating expansion, and they all satisfy two assumptions: $f(0) = 0$ and $f(R) \to R - 2\Lambda$ as $R \gg R_c$ [46]. Here $f(R) = R - 2\Lambda$ is the traditional $\Lambda$CDM model, $\Lambda$ is the effective cosmological constant, and $\Lambda$ is considered to be unrelated to the quantum vacuum energy in above models [46]. All kinds of constraints will be considered for these $f(R)$ models in Section 4.

In the following part, we are going to show a series of substitutions and calculation results to simplify the subsequent processes. Let us make the following definitions:

$$x_d \equiv \frac{R_d}{R_c}, \quad (24)$$

$$x \equiv \frac{R}{R_c}, \quad (25)$$

$$g_i(x) \equiv \frac{f_i(R)}{R_c} = \frac{f_i(R_c x)}{R_c}, \quad (26)$$

where $i = 1, 2, 3$, and the following stability conditions will almost be based on $x_d$. According to Eq. (25) and Eq. (26), ones have:

$$f_i'(R) = \frac{1}{R_c} f_i'(R_c x) = g_i'(x), \quad (27)$$

$$f_i''(R) = \frac{1}{R_c^2} f_i''(R_c x) = \frac{1}{R_c} g_i''(x), \quad (28)$$

$$f_i'''(R) = \frac{1}{R_c^3} f_i'''(R_c x) = \frac{1}{R_c^2} g_i'''(x), \quad (29)$$

$$f_i''''(R) = \frac{1}{R_c^4} f_i''''(R_c x) = \frac{1}{R_c^3} g_i''''(x), \quad (30)$$

where $f_i'(R) = df_i(R)/dR$, $f_i'(R_c x) = df_i(R_c x)/dx$, $g_i'(x) = dg_i(x)/dx$ and the same goes for higher derivatives. An agreement is made in our follow-up parts: the object of derivation depends on what the variable is in the parentheses behind the function. Besides, when the variable has a subscript '$d$', it means taking the derivative of the variable at de Sitter point such as $g_i'(x_d) = g_i'(x)|_{x=x_d}$ (here $x_d$ is also

changeable for different de Sitter points). Bring Eqs. (21), (22) and (23) into Eq. (26) to individually get:

$$g_1(x) = x - \alpha_1 \frac{x^{2\beta_1}}{x^{2\beta_1} + 1}, \tag{31}$$

$$g_2(x) = x - \alpha_2[1 - (1 + x^2)^{-\beta_2}], \tag{32}$$

$$g_3(x) = x - \frac{\alpha_3}{\pi}\operatorname{arccot}(x^{-2}) - \beta_3(1 - e^{-x}). \tag{33}$$

And Eqs. (3) and (4) can be updated by Eqs. (24), (26), (27) and (28) to:

$$m_i^2 = \frac{R_c}{3}\left(\frac{g_i'(x_d)}{g_i''(x_d)} - x_d\right), \tag{34}$$

$$g_i'(x_d)x_d - 2g_i(x_d) = 0. \tag{35}$$

## 3. Constraints for stable massless scalar polarization

In this section, we will show the detailed constraint conditions for $f(R)$ models and for their stable massless scalar polarizations. These constraints mainly come from the cosmology, the massless scalar state and the stability. We are going to introduce them from two aspects: one is the cosmology, and the other one is the effective potential that can not only be used to reflect a massless scalar state but also to describe a stable perturbation of space-time.

### 3.1 Constraints from cosmology

For $R \geq R_0 > 0$, the following stability conditions for $f(R)$ models must be met:

$$f'(R) > 0, \tag{36}$$

$$f''(R) > 0, \tag{37}$$

where $R_0$ stands for the Ricci scalar in the infinite future [46]. The first condition Eq. (36) ensures that the graviton is not a ghost [47]. If Eq. (36) is violated, the homogeneity and isotropy of the Universe would be lost [48, 49]. The second condition Eq. (37) is provided to avoid the Dolgov-Kawasaki instability [47, 50]. Besides, Eq. (37) was also considered to prevent the scalaron from being a tachyon or a ghost [46]. Employing Eqs. (27) and (28), we can turn Eqs. (36) and (37) into:

$$g_i'(x) > 0, \tag{38}$$

$$g_i''(x) > 0, \tag{39}$$

where $0 < x_0 \leq x$ and $x_0 \equiv R_0/R_c$. For the de Sitter curvature, we have $x_d \in [x_0, +\infty)$, and hence the above inequations lead to:

$$g_i'(x_d) > 0, \quad (40)$$

$$g_i''(x_d) > 0. \quad (41)$$

What needs to be distinguished is: Eqs. (40) and (41) will be applied first to check the stability at the curvature point, and if they are met, Eqs. (38) and (39) will be applied later to check the stability in the curvature range.

In addition to the basic settings $\alpha_i, \beta_i > 0$ ($i = 1, 2, 3$), the parameter constraints of $f_i(R)$ from cosmology are provided by:

$$\beta_{1,2} > 0.9 \quad (42)$$

and

$$0 < \alpha_3 < 0.7. \quad (43)$$

Eq. (42) is given by considering the violations of the weak and strong equivalence principles, whose constraint range is stricter than the solar system constraints $\beta_{1,2} > 0.5$ [51]. Eq. (43) is given by using the gravitational wave event GW170817 [41, 52]. Besides, the above results are based on the first-order perturbations of the equations of motion, and the linear equations of motion will be naturally obtained when Eq. (4) (or Eq. (35)) is maintained to ensure a static de Sitter stage. The parameter $\alpha_i$ can be expressed by $x_d$ and $\beta_i$ according to Eq. (35). Substitute Eqs. (31), (32) and (33) into Eq. (35) to gain:

$$\alpha_1 = \frac{\frac{x_d}{2}}{\frac{x_d^{2\beta_1}}{x_d^{2\beta_1} + 1} - \beta_1 \frac{x_d^{2\beta_1}}{(x_d^{2\beta_1} + 1)^2}}, \quad (44)$$

$$\alpha_2 = \frac{\frac{x_d}{2}}{1 - (1 + x_d^2)^{-\beta_2} - \beta_2 x_d^2 (1 + x_d^2)^{-\beta_2 - 1}}, \quad (45)$$

$$\alpha_3 = \frac{\pi e^{-x_d}(1 + x_d^4)(x_d e^{x_d} + x_d \beta_3 - 2\beta_3 e^{x_d} + 2\beta_3)}{2(x_d^4 \mathrm{arccot}(x_d^{-2}) - x_d^2 + \mathrm{arccot}(x_d^{-2}))}. \quad (46)$$

It is important to point out that $\alpha_i(x_d)$ are just the values determined by $x_d$ rather than the functions of $x$, and $x_d$ in $\alpha_i(x_d)$ cannot be treated as a variable in the process of taking the all-order derivatives of $g_i(x)$ at $x_d$ (namely for: $g_i'(x_d)$, $g_i''(x_d)$, $g_i'''(x_d)$, $g_i''''(x_d) \cdots$). Moreover, Eq. (35) will be indirectly satisfied when Eqs. (44), (45) and (46) are used, and it means that using them will remove the need of considering Eq. (35).

## 3.2 Constraints from effective potential

Following [36, 53], we are allowed to define a scalar field $\Phi$ and an effective potential $V(\Phi)$ as follows:

$$\Phi \equiv f'(R), \tag{47}$$

$$V(\Phi) \equiv \frac{1}{3}\int \left(2f(R) - f'(R)R\right) d\Phi. \tag{48}$$

The above definitions make the trace of the field equation turn into a Klein-Gordon equation for the scalar field, namely: $f'(R)R + 3\Box f'(R) - 2f(R) = 0 \to \Box \Phi = dV(\Phi)/d\Phi$. Based on the Klein-Gordon equation, in order to keep a stable perturbation of space-time, the background scalar $\Phi_d$ is strongly required to stay at a local minimum of the effective potential $V(\Phi)$ [36, 54]. The most direct and simple measures are:

$$V'(\Phi_d) = 0, \tag{49}$$

$$V''(\Phi_d) > 0, \tag{50}$$

where $\Phi_d \equiv f'(R_d)$ and Eq. (49) is actually equivalent to Eq. (4). Due to Eq. (3), one has $V''(\Phi_d) = m^2$, so Eq. (50) means $m^2 > 0$, agreeing with most suggestions [28, 35, 36]. As a result, the massless scalar polarization of $f(R)$ gravity seems to be excluded.

However, Eqs. (49) and (50) are overly strict conditions for keeping $\Phi_d$ at a local minimum. For some specific $f(R)$ models, this purpose is also able to be reached in the case of $m = 0$ [37]. In [55], the higher derivative test provides a mathematical result that a function of one real variable $y(x)$ has a local minimum point $(x_0, y(x_0))$ when $y'(x_0) = y''(x_0) = \cdots = y^{(2n-1)}(x_0) = 0$ and $y^{(2n)}(x_0) > 0$ ($n$ is a positive whole number). Therefore, we can give more general constraints for building a local minimum in the effective potential:

$$V^{(1)}(\Phi_d) = V^{(2)}(\Phi_d) = \cdots = V^{(2K-1)}(\Phi_d) = 0, \tag{51}$$

$$V^{(2K)}(\Phi_d) > 0, \tag{52}$$

where $K = 1, 2, 3 \cdots$, the superscript $(N)$ stands for the $N$th derivative of $V(\Phi)$ at $\Phi = \Phi_d$, and the derivatives of $V(\Phi)$ are assumed to be continuous around $\Phi_d$. The existence of a local minimum needs that the constraints Eqs. (51) and (52) hold for at least one value of $K$, while the existence of a stable massless scalar polarization requires that the constraints Eqs. (51) and (52) hold for at least one value of $K \geq 2$. For $K = 1$, Eqs. (51) and (52) can return to the original constraints Eqs. (49) and (50).

For $K \geq 2$, if these constraints in Eq. (51) are independent of each other, there seem to be more constraints for $f(R)$ to meet with the increase of $K$, so that more free parameters would need to be included in $f(R)$. However, if these constraints are dependent, the constraints of $K_1$ ($K_1 > K_2$) might not be stricter than those of $K_2$. Therefore, every case of $K \geq 2$ is supposed to be considered for stable massless scalar polarizations. To start with, the constraints of the effective potential in the case of $K = 2$ will be discussed:

$$V_i'(\Phi_d) = V_i''(\Phi_d) = V_i'''(\Phi_d) = 0, \tag{53}$$

$$V_i''''(\Phi_d) > 0, \tag{54}$$

where the subscript $i$ is added to represent different effective potentials coming from $f_i(R)$. When the researched $f(R)$ models could not survive in the case of $K = 2$, the case of $K \geq 3$ will be discussed for them later. Take Eqs. (25), (26) and (27) into Eqs. (47) and (48) to obtain:

$$d\Phi = g_i''(x)dx, \tag{55}$$

$$V_i(\Phi) = \frac{1}{3}\int R_c(2g_i(x) - g_i'(x)x)\,d\Phi. \tag{56}$$

At $x = x_d$, the above equations lead to:

$$V_i'(\Phi_d) = \frac{R_c}{3}(2g_i(x_d) - g_i'(x_d)x_d), \tag{57}$$

$$V_i''(\Phi_d) = \frac{R_c}{3}\left(\frac{g_i'(x_d)}{g_i''(x_d)} - x_d\right), \tag{58}$$

$$V_i'''(\Phi_d) = \frac{R_c}{3}\frac{g_i'(x_d)(-g_i'''(x_d))}{(g_i''(x_d))^3}, \tag{59}$$

$$V_i''''(\Phi_d) = \frac{R_c}{3}\frac{1}{(g_i''(x_d))^5}\Big(3g_i'(x_d)(g_i'''(x_d))^2 - g_i'(x_d)g_i''(x_d)g_i''''(x_d)$$

$$-(g_i''(x_d))^2 g_i'''(x_d)\Big). \tag{60}$$

By bringing Eqs. (57), (58) (59) and (60) into the constraints of the effective potential Eqs. (53) and (54) and considering Eqs. (34), (40) and (41), we have:

$$V_i'(\Phi_d) = 0 \Rightarrow 2g_i(x_d) - g_i'(x_d)x_d = 0, \tag{61}$$

$$V_i''(\Phi_d) = 0 \Rightarrow m_i^2 = \frac{R_c}{3}\left(\frac{g_i'(x_d)}{g_i''(x_d)} - x_d\right) = 0, \tag{62}$$

$$V_i'''(\Phi_d) = 0 \Rightarrow g_i'''(x_d) = 0, \tag{63}$$

$$V_i''''(\Phi_d) > 0 \Rightarrow g_i''''(x_d) < 0, \tag{64}$$

where Eq. (64) is obtained under the premise $g_i'''(x_d) = 0$. In addition to ensuring a local minimum, the above constraints Eqs. (61) and (62) also overlap with some other requirements: Eq. (61) indicates that the de Sitter point is a stationary point, corresponding to Eq. (35); Eq. (62) is the massless scalar condition for $f(R)$, corresponding to Eq. (34) $= 0$. Moreover, the above constraints can be converted to the forms described by $f(R_d)$ if Eqs. (24), (26), (27), (28), (29) and (30) are used.

## 4. Retesting $f(R)$ gravity with the updated constraints

First, the special limiting point $x_d = 0^+$ should be discussed, because the de Sitter point might become a local minimum point at $x_d = 0^+$, even though not all constraints in Eqs. (51) and (52) are satisfied. When $x_d = 0^+$, the stability condition Eq. (41) would be violated: $g_1''(0^+) = 2(0^+)^{-2+2\beta_1}\alpha_1\beta_1(1 - 2\beta_1) < 0$ ($\beta_1 > 0.9$), $g_2''(0^+) = -2\alpha_2\beta_2 < 0$ and $g_3''(0^+) = -2\alpha_3/\pi + \beta_3 < 0$ ($\alpha_3 \to +\infty$ in Eq. (46) as $x_d = 0^+$). For these reasons, the special limiting point $x_d = 0^+$ is an impracticable point and needs to be eliminated.

In this section, the constraints provided by Section 3 will be applied to the specific $f(R)$ models including Hu-Sawicki Eq. (31), Starobinsky Eq. (32) and Gogoi-Dev Eq. (33). The parameter spaces of $x_d$ and $\beta_i$ are built, in which two viable areas are defined: one is a light orange area (active area) described by the constraints of the cosmology Eqs. (40), (41), (42) and (43), while the other one is a light red area described by the constraints of the effective potential Eq. (64). Besides, two lines are defined: one is a blue solid line presenting the solutions of $m_i^2 = 0$, and the other one is a brown solid line presenting the solutions of $g_i'''(x_d) = 0$. Only when the intersection of the two lines occurs in both the colored areas at the same time, will all the constraints for $K = 2$ be met, so that the corresponding $f(R)$ models can reflect a stable massless scalar polarization.

### 4.1 Active local minimum points

In figure 1, a possible intersection of $m_1^2 = 0$ and $g_1'''(x_d) = 0$ lies within the light red area but is clearly outside the light orange area. It implies that the constraints cannot be met concurrently, and hence the Hu-Sawicki model is inapplicable for stable massless scalar polarizations in the case of $K = 2$. In figure 2, it can be seen that the light orange area wraps $m_2^2 = 0$ and $g_2'''(x_d) = 0$ while the light red area wraps $g_2'''(x_d) = 0$. Because there is no intersection between the two lines (even for $\beta_2 \gg$

1), the Starobinsky model is also inapplicable for stable massless scalar polarizations in the case of $K = 2$.

Considering the cases of $K \geq 3$ for Hu-Sawicki and Starobinsky models, they have common requirements: $V_i''(\Phi_d) = V_i'''(\Phi_d) = 0$, which means that the two lines $m_i^2 = 0$ and $g_i'''(x_d) = 0$ should intersect in parameter space. In figure 1, we can see that the only possible intersection of Hu-Sawicki is around $\beta_1 = 0.5$, but Eq. (42) requires $\beta_1 > 0.9$. Thus, there is no feasible intersection for Hu-Sawicki. In figure 2, we can see that the two lines for Starobinsky do not have any intersection. To sum up, none of cases of $K \geq 2$ can hold in the models of Hu-Sawicki and Starobinsky, so the two models fail to build stable massless scalar polarizations.

In figure 3, there is an intersection between $m_3^2 = 0$ and $g_3'''(x_d) = 0$, and its coordinate is around ($\beta_3 = 1.0441$, $x_d = 0.9077$). By using Eq. (46), one obtains $\alpha_3 = 0.3511$, agreeing with the required range $0 < \alpha_3 < 0.7$. Different from figure 1 and figure 2, the intersection occurs in both the light red and the light orange areas at the same time, which means that all constraint conditions can be met and an active local minimum point can be reached. Therefore, Gogoi-Dev model is feasible for stable massless scalar polarizations.

The results indicate that the models of Hu-Sawicki and Starobinsky cannot be used to build stable massless scalar polarizations, while Gogoi-Dev model can do that. Because these involved models have the same freedom degree, the reason for the difference should lie in the model structure rather than the number of free parameters. Obvious evidence is that the colored areas in figures 1, 2 and 3 visibly own different shapes, and these shapes are drawn by the derivatives and parameter bounds of $f(R)$. In the next subsection, we are going to examine the feasibility of the massless case of Gogoi-Dev model in more detail.

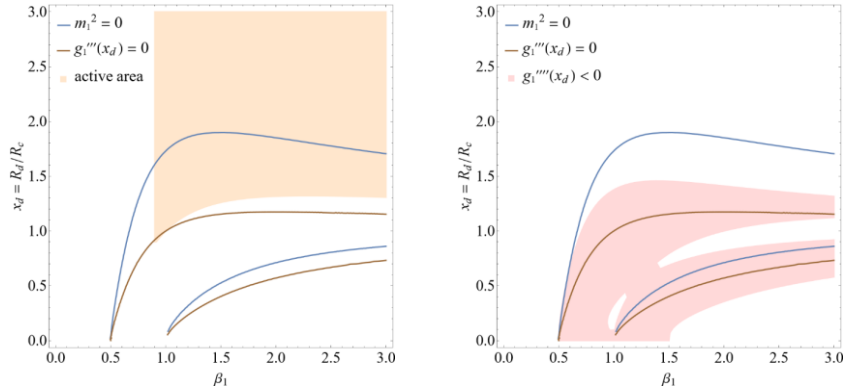

**Figure 1.** The parameter space of $x_d$ and $\beta_1$ is used for Hu-Sawicki model Eq. (31), where $\alpha_1$

in Eq. (31) has been replaced by Eq. (44). The blue solid line presents the solutions of $m_1^2 = 0$, and the brown solid line presents the solutions of $g_1'''(x_d) = 0$; The light orange area (active area) is given by $g_1'(x_d) > 0$, $g_1''(x_d) > 0$, $\alpha_1 > 0$ and $\beta_1 > 0.9$, and the light red area is given by $g_1''''(x_d) < 0$.

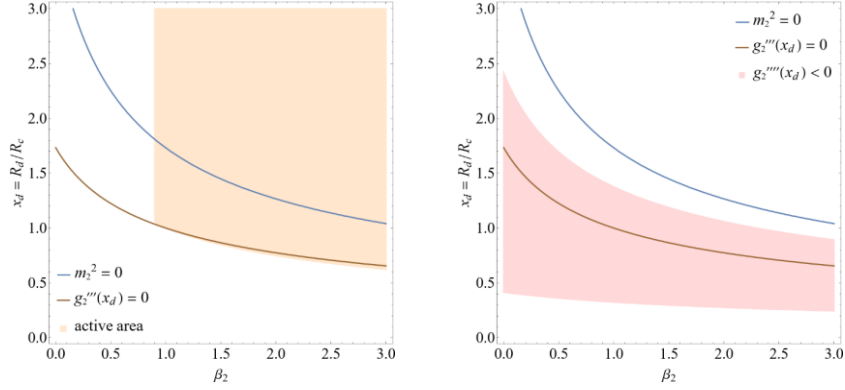

**Figure 2.** The parameter space of $x_d$ and $\beta_2$ is used for Starobinsky model Eq. (32), where $\alpha_2$ in Eq. (32) has been replaced by Eq. (45). The blue solid line presents the solutions of $m_2^2 = 0$, and the brown solid line presents the solutions of $g_2'''(x_d) = 0$; The light orange area (active area) is given by $g_2'(x_d) > 0$, $g_2''(x_d) > 0$, $\alpha_2 > 0$ and $\beta_2 > 0.9$, and the light red area is given by $g_2''''(x_d) < 0$.

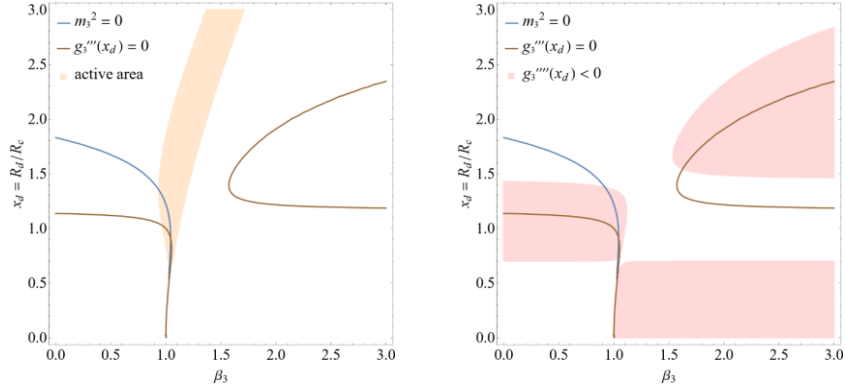

**Figure 3.** The parameter space of $x_d$ and $\beta_3$ is used for Gogoi-Dev model Eq. (33), where $\alpha_3$ in Eq. (33) has been replaced by Eq. (46). The blue solid line presents the solutions of $m_3^2 = 0$, and the brown solid line presents the solutions of $g_3'''(x_d) = 0$; The light orange area (active area) is given by $g_3'(x_d) > 0$, $g_3''(x_d) > 0$, $0 < \alpha_3 < 0.7$ and $\beta_3 > 0$, and the light red area is given by $g_2''''(x_d) < 0$.

### 4.2 Satisfied constraint conditions

The constraint conditions Eqs. (61), (62), (63) and (64) are aimed at giving a local minimum of the effective potential $V(\Phi)$. In order to show this clearly, some figures will be drawn for Gogoi-Dev model to reflect how $V_3(\Phi)$ changes with $\Phi$. First, based on the intersection coordinate given in Subsection 4.1, the model parameters $\alpha_3$ and $\beta_3$ can be determined. After that, we apply Eqs. (55), (56) and (57) to convert $\Phi$, $V_3{'}(\Phi)$ and $V_3(\Phi)$ into the functions of $x$, so that the figures about $V_3{'}(\Phi)$ and $V_3(\Phi)$ versus $\Phi$ can be drawn with the change of $x$. In figure 4, there is a zero value for $V_3{'}(\Phi)$ at $\Phi_d = 0.4579$. And $V_3{'}(\Phi) < 0$ on the left-hand side of $\Phi_d$, while $V_3{'}(\Phi) > 0$ on the right-hand side of $\Phi_d$. Thus, $V_3(\Phi)$ has a local minimum at $\Phi_d$, agreeing with the result of the picture of $V_3(\Phi)/R_c$. For avoiding the effect of extreme point, $V_3(\Phi)/R_c$ is drawn by letting the integral range of Eq. (56) be from 0.8 to $x$ ($0.8 \leq x \leq 1.0$).

Eqs. (40) and (41), the stability conditions for $f(R)$ at $x_d$, have been satisfied for Gogoi-Dev model. Therefore, Eqs. (38) and (39), the stability conditions for $f(R)$ in $[x_0, +\infty)$, should be checked in the following part. Because $0 < x_0 \leq x_d$, both $g_3{'}(x) > 0$ and $g_3{''}(x) > 0$ need to be met in $[x_0, x_d]$ and $[x_d, +\infty)$. The range of $[x_0, x_d]$ will not be strictly defined, since it depends on the different values of the model parameter $R_c$. In other words, if there is a continuous range tightly attached to the left-hand side of $x_d$, and $g_i{'}(x) > 0$ and $g_i{''}(x) > 0$ are met in this continuous range, we believe that these stability conditions could be satisfied in $[x_0, x_d]$. In figure 5, we can see that $g_3{''}(x) > 0$ in $[0, +\infty)$, so the stability condition Eq. (39) is true for Gogoi-Dev model. On the other hand, there are $g_3{'}(x) > 0$ in $[h, x_d]$ ($0 < h < x_d$) and $[x_d, +\infty)$. Because $[h, x_d]$ can be regarded as $[x_0, x_d]$ with the adjustable parameter $R_c$, the stability condition Eq. (38) is also true for Gogoi-Dev model. To sum up, the stability in the curvature range is not violated for Gogoi-Dev model.

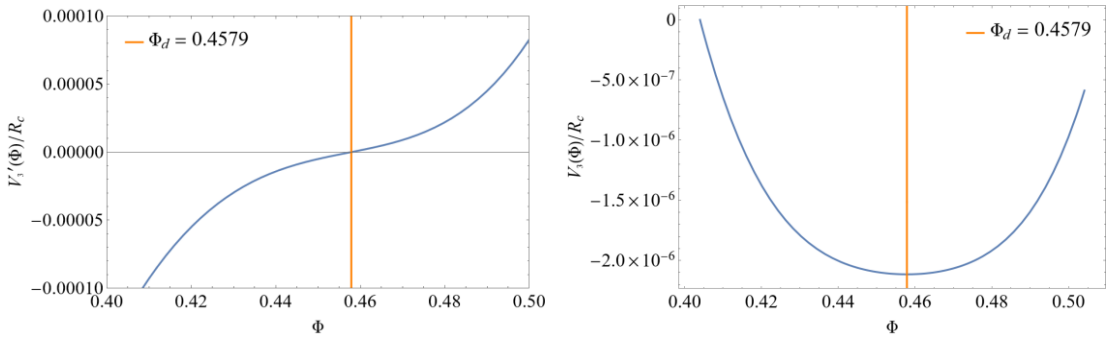

**Figure 4.** $V_3{'}(\Phi)/R_c$ and $V_3(\Phi)/R_c$ versus $\Phi$ in the region of $x = 0.8$ to $1.0$ for Gogoi-Dev model Eq. (33) with $\alpha_3 = 0.3511$ and $\beta_3 = 1.0441$, where the orange solid line stands for the value of scalar field at de Sitter point $\Phi_d = 0.4579$.

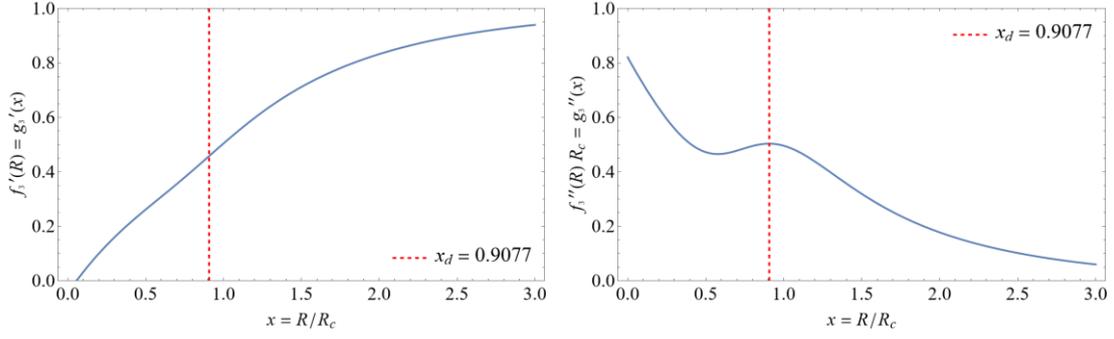

**Figure 5.** $g_3'(x)$ and $g_3''(x)$ versus $x$ in the region of $x = 0.0$ to $3.0$ for Gogoi-Dev model Eq. (33) with $\alpha_3 = 0.3511$ and $\beta_3 = 1.0441$, where the red dotted line stands for the location of de Sitter point $x_d = 0.9077$.

## 5. Discussions and conclusions

The additional polarization of $f(R)$ gravity is a scalar polarization, which is a combination of the longitudinal and the breathing modes. When the scalar mass given by Eq. (3) vanishes, the scalar polarization will reduce to a pure breathing mode. However, to maintain a stable perturbation of space-time, the constraints of the effective potential in Eqs. (49) and (50) require the scalar mass to always be positive, seemingly excluding the massless scalar polarization in $f(R)$ gravity. But we indicate that the constraints should be adjusted to more general forms Eqs. (51) and (52), and a stable perturbation needs the constraints Eqs. (51) and (52) hold for at least one value of $K$. The original constraints are just the special case of $K = 1$, while the cases of $K \geq 2$ can ensure that both stable perturbation and massless scalar state are built.

To get a stable massless scalar polarization in $f(R)$ gravity, the constraints from two aspects are considered: one is the cosmology in Subsection 3.1, and the other one is the effective potential in Subsection 3.2. For the 3-parameter $f(R)$ models in Eqs. (21), (22) and (23), which are regarded as viable dark energy models, we have analyzed the possibility of the existence of stable massless scalar polarizations. The results show that both Hu-Sawicki and Starobinsky models fail to build stable massless scalar polarizations. On the contrary, Gogoi-Dev model can meet all kinds of constraints and lead to a stable massless scalar polarization. Therefore, the existence of stable massless scalar polarizations in $f(R)$ gravity should not be ignored. In other words, if a massless scalar polarization (or pure breathing mode) is observed, $f(R)$ gravity should not be removed from those alternative modified gravity models.

All the tested $f(R)$ models have 3 parameter degrees of freedom, but their abilities to maintain stable massless scalar polarizations vary significantly. It means that the model structure (or function expression) of $f(R)$ gravity could influence the results of scalar polarization. That is to say, the scalar polarization is model-dependent for $f(R)$ gravity. It is possible to distinguish between various $f(R)$ models by examining whether they have stable massless scalar polarizations. On the other hand, we can also apply stable massless scalar polarizations to provide some observation constraints on the free parameters of $f(R)$ models. Furthermore, we are allowed to construct $f(R)$ models based on the stable massless scalar polarization. The possible prescription under the metric formalism and the first-order linear perturbation, is to make the established $f(R)$ models meet the cosmology constraints, and the potential constraints Eqs. (51) and (52) for at least one value of $K \geq 2$.

Finally, some feasible research directions should be mentioned. In [56], various polarization modes of gravitational waves are studied in higher-order gravity, which involves more general scalar invariants besides the Ricci scalar. Our results are expected to be applied to this model to give more general analysis. In [57], the impact of the chameleon mechanism on observations is discussed in the framework of $f(R)$ gravity. It is interesting to explore the detectability of the massless scalar waves according to the detection environment. In [19], a special evolution of gravitons can be driven by $f(R)$ cosmological model. After imposing parameter constraints of stable massless scalar polarization, the evolution of gravitons deserves further study.

## Acknowledgments

We are grateful to Bao-Min Gu for helpful discussions. The work is in part supported by NSFC Grant No.12205104, "the Fundamental Research Funds for the Central Universities" with Grant No. 2023ZYGXZR079, the Guangzhou Science and Technology Project with Grant No. 2023A04J0651 and the startup funding of South China University of Technology.

## References

[1] A.G. Riess, et al., Observational Evidence from Supernovae for an Accelerating Universe and a Cosmological Constant, The Astronomical Journal, 116 (1998) 1009.
[2] M. Tegmark, et al., Cosmological parameters from SDSS and WMAP, Physical Review D, 69 (2004) 103501.

[3] D.N. Spergel, et al., Three-Year Wilkinson Microwave Anisotropy Probe (WMAP) Observations: Implications for Cosmology, The Astrophysical Journal Supplement Series, 170 (2007) 377.

[4] E.J. Copeland, M. Sami, S. Tsujikawa, DYNAMICS OF DARK ENERGY, International Journal of Modern Physics D, 15 (2006) 1753-1935.

[5] K. Bamba, S. Capozziello, S.i. Nojiri, S.D. Odintsov, Dark energy cosmology: the equivalent description via different theoretical models and cosmography tests, Astrophysics and Space Science, 342 (2012) 155-228.

[6] T.P. Sotiriou, V. Faraoni, f(R) theories of gravity, Reviews of Modern Physics, 82 (2010) 451-497.

[7] A. De Felice, S. Tsujikawa, f(R) Theories, Living Reviews in Relativity, 13 (2010) 3.

[8] C. Brans, R.H. Dicke, Mach's Principle and a Relativistic Theory of Gravitation, Physical Review, 124 (1961) 925-935.

[9] G.W. Horndeski, Second-order scalar-tensor field equations in a four-dimensional space, International Journal of Theoretical Physics, 10 (1974) 363-384.

[10] R.V. Wagoner, Scalar-Tensor Theory and Gravitational Waves, Physical Review D, 1 (1970) 3209-3216.

[11] S.i. Nojiri, S.D. Odintsov, Unified cosmic history in modified gravity: From F(R) theory to Lorentz non-invariant models, Physics Reports, 505 (2011) 59-144.

[12] S. Nojiri, S.D. Odintsov, V.K. Oikonomou, Modified gravity theories on a nutshell: Inflation, bounce and late-time evolution, Physics Reports, 692 (2017) 1-104.

[13] A.S. Sefiedgar, K. Atazadeh, H.R. Sepangi, Generalized virial theorem in Palatini f(R) gravity, Physical Review D, 80 (2009) 064010.

[14] L.S. Collaboration, et al., Observation of Gravitational Waves from a Binary Black Hole Merger, Physical Review Letters, 116 (2016) 061102.

[15] L.S. Collaboration, et al., GW170817: Observation of Gravitational Waves from a Binary Neutron Star Inspiral, Physical Review Letters, 119 (2017) 161101.

[16] J. Luo, et al., TianQin: a space-borne gravitational wave detector, Classical and Quantum Gravity, 33 (2016) 035010.

[17] G. Wang, W.-B. Han, Observing gravitational wave polarizations with the LISA-TAIJI network, Physical Review D, 103 (2021) 064021.

[18] T. Liu, X. Lou, J. Ren, Pulsar Polarization Arrays, Physical Review Letters, 130 (2023) 121401.

[19] S. Capozziello, M. De Laurentis, S. Nojiri, S.D. Odintsov, Evolution of gravitons in accelerating cosmologies: The case of extended gravity, Physical Review D, 95 (2017) 083524.

[20] S.i. Nojiri, S.D. Odintsov, Cosmological bound from the neutron star merger GW170817 in scalar–tensor and F(R) gravity theories, Physics Letters B, 779 (2018) 425-429.

[21] S.D. Odintsov, V.K. Oikonomou, F.P. Fronimos, Quantitative predictions for f(R) gravity primordial gravitational waves, Physics of the Dark Universe, 35 (2022) 100950.

[22] D.M. Eardley, D.L. Lee, A.P. Lightman, Gravitational-Wave Observations as a Tool for Testing Relativistic Gravity, Physical Review D, 8 (1973) 3308-3321.

[23] L.S.C. The, et al., Tests of general relativity with the binary black hole signals from the LIGO-Virgo catalog GWTC-1, Physical Review D, 100 (2019) 104036.

[24] P.T.H. Pang, R.K.L. Lo, I.C.F. Wong, T.G.F. Li, C. Van Den Broeck, Generic searches for alternative gravitational wave polarizations with networks of interferometric detectors, Physical Review D, 101 (2020) 104055.

[25] K.J. Lee, F.A. Jenet, R.H. Price, Pulsar Timing as a Probe of Non-Einsteinian Polarizations of Gravitational Waves, The Astrophysical Journal, 685 (2008) 1304.

[26] R.C. Bernardo, K.-W. Ng, Constraining gravitational wave propagation using pulsar timing array correlations, Physical Review D, 107 (2023) L101502.

[27] F.A.E. Pirani, Republication of: On the physical significance of the Riemann tensor, General Relativity and Gravitation, 41 (2009) 1215-1232.

[28] D. Liang, Y. Gong, S. Hou, Y. Liu, Polarizations of gravitational waves in f(R) gravity, Physical Review D, 95 (2017) 104034.

[29] M.E.S. Alves, O.D. Miranda, J.C.N. de Araujo, Probing the f(R) formalism through gravitational wave polarizations, Physics Letters B, 679 (2009) 401-406.

[30] Y.-H. Hyun, Y. Kim, S. Lee, Exact amplitudes of six polarization modes for gravitational waves, Physical Review D, 99 (2019) 124002.

[31] É.É. Flanagan, S.A. Hughes, The basics of gravitational wave theory, New Journal of Physics, 7 (2005) 204.

[32] Y.-Q. Dong, Y.-Q. Liu, Y.-X. Liu, Polarization modes of gravitational waves in general modified gravity: General metric theory and general scalar-tensor theory, Physical Review D, 109 (2024) 044013.

[33] M.E.S. Alves, Testing gravity with gauge-invariant polarization states of gravitational waves: Theory and pulsar timing sensitivity, Physical Review D, 109

(2024) 104054.

[34] D.J. Gogoi, U. Dev Goswami, Gravitational waves in f(R) gravity power law model, Indian Journal of Physics, 96 (2022) 637-646.

[35] S. Tsujikawa, Observational signatures of f(R) dark energy models that satisfy cosmological and local gravity constraints, Physical Review D, 77 (2008) 023507.

[36] Y. Louis, L. Chung-Chi, G. Chao-Qiang, Gravitational waves in viable f(R) models, Journal of Cosmology and Astroparticle Physics, 2011 (2011) 029.

[37] V. Müller, H.J. Schmidt, A.A. Starobinsky, The stability of the de Sitter space-time in fourth order gravity, Physics Letters B, 202 (1988) 198-200.

[38] X. Zhang, J. Yu, T. Liu, W. Zhao, A. Wang, Testing Brans-Dicke gravity using the Einstein telescope, Physical Review D, 95 (2017) 124008.

[39] T. Faulkner, M. Tegmark, E.F. Bunn, Y. Mao, Constraining f(R) gravity as a scalar-tensor theory, Physical Review D, 76 (2007) 063505.

[40] J. Lu, Y. Wu, W. Yang, M. Liu, X. Zhao, The generalized Brans-Dicke theory and its cosmology, The European Physical Journal Plus, 134 (2019) 318.

[41] D.J. Gogoi, U. Dev Goswami, A new f(R) gravity model and properties of gravitational waves in it, The European Physical Journal C, 80 (2020) 1101.

[42] J.D. Barrow, A.C. Ottewill, The stability of general relativistic cosmological theory, Journal of Physics A: Mathematical and General, 16 (1983) 2757.

[43] Y. Gong, S. Hou, The Polarizations of Gravitational Waves, in: Universe, 2018.

[44] G. Cognola, et al., Class of viable modified f(R) gravities describing inflation and the onset of accelerated expansion, Physical Review D, 77 (2008) 046009.

[45] W. Hu, I. Sawicki, Models of f(R) cosmic acceleration that evade solar system tests, Physical Review D, 76 (2007) 064004.

[46] A.A. Starobinsky, Disappearing cosmological constant in f(R) gravity, JETP Letters, 86 (2007) 157-163.

[47] S.A. Appleby, R.A. Battye, A.A. Starobinsky, Curing singularities in cosmological evolution of F(R) gravity, Journal of Cosmology and Astroparticle Physics, 2010 (2010) 005.

[48] H. Nariai, Gravitational Instability of Regular Model-Universes in a Modified Theory of General Relativity, Progress of Theoretical Physics, 49 (1973) 165-180.

[49] V.T. Gurovich, A.A. Starobinskii, Quantum effects and regular cosmological models, Zh. Eksp. Teor. Fiz. (USSR), 77 (1979) 1683-1700.

[50] A.D. Dolgov, M. Kawasaki, Can modified gravity explain accelerated cosmic expansion?, Physics Letters B, 573 (2003) 1-4.

[51] S. Capozziello, S. Tsujikawa, Solar system and equivalence principle constraints on f(R) gravity by the chameleon approach, Physical Review D, 77 (2008) 107501.

[52] S. Jana, S. Mohanty, Constraints on f(R) theories of gravity from GW170817, Physical Review D, 99 (2019) 044056.

[53] S. Capozziello, A. Stabile, A. Troisi, Spherically symmetric solutions in f(R) gravity via the Noether symmetry approach, Classical and Quantum Gravity, 24 (2007) 2153.

[54] M.M. Ivanov, A.V. Toporensky, STABLE SUPER-INFLATING COSMOLOGICAL SOLUTIONS IN f(R)-GRAVITY, International Journal of Modern Physics D, 21 (2012) 1250051.

[55] S.W. Nash, The Higher Derivative Test for Extrema and Points of Inflection, The American Mathematical Monthly, 66 (1959) 709-713.

[56] C. Bogdanos, S. Capozziello, M.D. Laurentis, S. Nesseris, Massive, massless and ghost modes of gravitational waves from higher-order gravity, Astroparticle Physics, 34 (2010) 236-244.

[57] T. Katsuragawa, T. Nakamura, T. Ikeda, S. Capozziello, Gravitational waves in F(R) gravity: Scalar waves and the chameleon mechanism, Physical Review D, 99 (2019) 124050.